\documentclass{sig-alternate}
\usepackage[letterpaper=true,pdfview=FitH,pdfstartview=FitH,bookmarks=true]{hyperref}
\pdfoutput=1

\usepackage{color}
\usepackage{algorithmic}
\usepackage{algorithm}
\usepackage{graphicx}     		
\usepackage{url}

\newcommand{\DB}[1]{\ensuremath{\mathcal{D}_{#1}}}
\newcommand{\Domain}[0]{\ensuremath{\mathcal{X}}}
\newcommand{\term}[1]{\emph{#1}}
\newcommand{\Reals}[0]{\ensuremath{\mathbb{R}}}
\newcommand{\Euclidean}[1]{\ensuremath{\Reals^{#1}}}
\newcommand{\TargetMatch}[0]{\ensuremath{\mathcal{R}}}
\newcommand{\PredictedMatch}[0]{\ensuremath{\hat{\mathcal{R}}}}

\newcommand{\PredictedMatchOM}[0]{\ensuremath{\PredictedMatch_{\mbox{\scriptsize 1-M}}}}
\newcommand{\PredictedMatchMO}[0]{\ensuremath{\PredictedMatch_{\mbox{\scriptsize M-1}}}}

\newcommand{\TruthPos}[0]{\ensuremath{\mathcal{R}^\star_+}}
\newcommand{\TruthNeg}[0]{\ensuremath{\mathcal{R}^\star_-}}
\newcommand{\AlgMatchMM}[0]{\textsc{ManyMany}}
\newcommand{\AlgMatchFC}[0]{\textsc{FirstChoice}}
\newcommand{\AlgMatchMFC}[0]{\textsc{MutualFirstChoice}}
\newcommand{\AlgMatchGr}[0]{\textsc{Greedy}}
\newcommand{\AlgMatchMW}[0]{\textsc{MaxWeight}}
\newcommand{\bigOh}[1]{\ensuremath{\mathcal{O}\left(#1\right)}}

\newcommand{\bigTheta}[1]{\ensuremath{\Theta\left(#1\right)}}
\renewcommand{\Pr}[1]{\ensuremath{\mathrm{Pr}\left(#1\right)}}

\renewcommand{\v}[1]{\ensuremath{\mathbf{#1}}}
\newcommand{\eg}[0]{\emph{e.g.},}
\newcommand{\etal}[0]{\emph{et al.},}
\newcommand{\ie}[0]{\emph{i.e.},}
\newcommand{\cf}[0]{\emph{cf.}}

\begin{document}

%

%

\title{Improving Entity Resolution with Global Constraints}
\subtitle{[Microsoft Research Technical Report MSR-TR-2011-100]}

%
%
%
%
%

\numberofauthors{3}
%

\author{
%
Jim~Gemmell \hspace{2em} Benjamin~I.~P.~Rubinstein \hspace{1.5em} Ashok~K.~Chandra \\[0.4em]
{\affaddr Microsoft Research, Mountain View, CA }\\[0.2em]
{\affaddr \{jgemmell, ben.rubinstein, achandra\}@microsoft.com}
}

\maketitle

\begin{abstract}
Some of the greatest advances in web search have come from leveraging socio-economic properties of online user behavior.  Past advances include PageRank, anchor text, hubs-authorities, and TF-IDF.  In this paper, we investigate another socio-economic property that, to our knowledge, has not yet been exploited: sites that create lists of entities, such as IMDB and Netflix, have an incentive to avoid gratuitous duplicates.  We leverage this property to resolve entities across the different web sites, and find that we can obtain substantial improvements in resolution accuracy.  This improvement in accuracy also translates into robustness, which often reduces the amount of training data that must be labeled for comparing entities across many sites.  Furthermore, the technique provides robustness when resolving sites that have some duplicates, even without first removing these duplicates. We present algorithms with very strong precision and recall, and show that max weight matching, while appearing to be a natural choice turns out to have poor performance in some situations.  The presented techniques are now being used in the back-end entity resolution system at a major Internet search engine.
\end{abstract}



\pagebreak

\section{Introduction} \label{sec:intro}

Some the greatest advancements in web search have come
from leveraging properties of the web that arise
due to social behavior or economic interests. PageRank uses
one such \term{socio-economic property} of the web that
reflects the preference of webmasters to link to popular pages,
and search engines use PageRank to figure out the most relevant
pages for any query~\cite{ilprints422}.  Anchor text is another socio-economic
property that represents the keywords by which the rest of the
world refers to a given page, even if these keywords are not
present in the page itself.  By exploiting these two socio-economic
properties together, Google made great advances over previous
search engines. Kleinberg~\cite{HubsAuthorities} utilized another socio-economic
property of the web: that for various topics there are people who
make lists of good web pages, and the bipartite graph of these hubs
(list pages) and authorities (linked pages) are mutually reinforcing.  

In this paper we investigate another online socio-economic property
that, to our knowledge, has never been exploited: 
that sites listing entities have an incentive 
to avoid gratuitous duplicates. For instance, 
duplicate movies in IMDB would have reviews and 
corrections applied to one copy and not the other. 
If Netflix has one entry for a DVD and a duplicate for the Blu-ray version, 
then their customers might be looking at one and not realize the other is available. 
Hulu supports Facebook likes for their movies and could have the like counts diluted by duplicates. 
Additional examples in other domains are easily constructed.
We leverage this socio-economic property to resolve
 entities across the different web sites, by applying a global one-to-one 
constraint to produced matchings.  The resulting resolution has much
 better accuracy compared to matching without such a constraint.  
Our framework for one-to-one entity resolution (ER) is generic in that it can
constrain existing resolution methods using weighted graph matching.
Our goal is not to engineer features or tune high-performance
domain-specific ER, but rather to develop generic algorithms
that can be combined with existing methods for improving
retrieval performance.

We perform a large-scale case-study in resolving movie entities from 
the complete catalogs of IMDB, Netix, iTunes and AMG, the largest of
 which contains roughly 500k entities. We use an active learning-based
 logistic regression score combiner with simple features and blocking,
 and find that constrained ER significantly outperforms unconstrained
 resolution, while being resilient to poor score tuning.
When compared with a crowd-sourced dataset of 
13K movie matches from Freebase, our algorithm improves on the 
existing high precision, and increases recall by a factor of over three.

\begin{figure}[t]
\centering
\includegraphics[width=7cm]{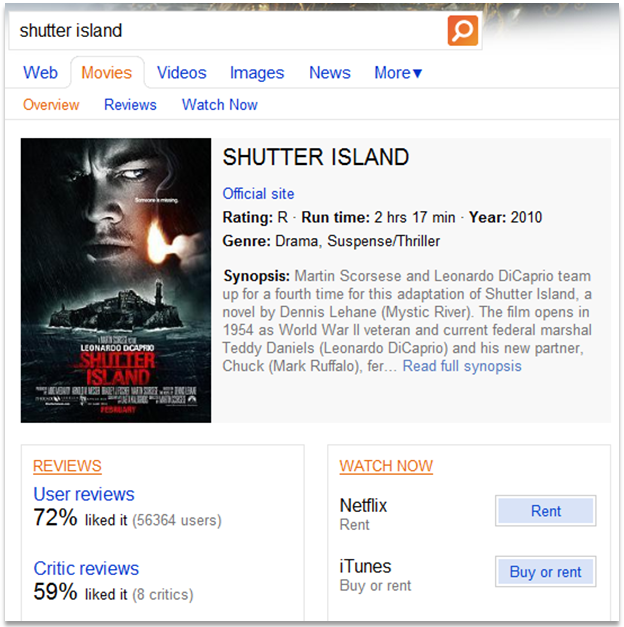}
\caption{Entity resolution is used in a major search engine's movies vertical for implementing ``buy'', ``rent'', ``watch'' entity actions and surfacing reviews.}
\label{fig:bing}
\end{figure}

The system presented here forms the basis for entity resolution as used
 in production in one of the major search engines on the Internet today. 
 The system aggregates movie information from a number of sources
 in order to support entity actions like
``rent", ``watch", and ``buy", as well as aggregating review
information (\cf\ Figure~\ref{fig:bing}). Because many sources need
to be merged, we aim to minimize the labeling and training
needed to incorporate new sources, in addition to requiring
excellent precision and recall.  It turns out that the one-to-one matching 
algorithms presented here are surprisingly robust, in that a scoring
 function for resolving entities across two sites often works effectively
 for matching other sites as well.


Finally we present a surprising result that demonstrates 
that the problem of graph matching for constrained 
entity resolution is fundamentally new---specifically 
that maximizing match weight does not necessarily 
optimize precision and recall---opening up a new direction 
for database and graph algorithm research. While 
exact max weight matching generally performs very well, it suffers 
extremely poor performance when very similar entities 
exist, resulting in both false negatives and false positives.


\subsection{Related Work}


Numerous previous works have studied entity resolution~\cite{InfoSys-MultClassSys,JCoopInfoSys,QDBMUD08-STEM,Benjelloun09,Kopcke2010}, statistical record linkage~\cite{JASA69,Winkler94,PAKDD08-FEBRL,Winkler06} and deduplication~\cite{ActiveLearn-KDD02,KDD03-MARLIN,Culotta2005,DedupSurvey07}. There has been almost no previous work producing one-to-one resolutions, and to our knowledge our work is the first to study the benefits of leveraging one-to-one structure in resolution problems, and the first to recognize the novel problem of graph matching for entity resolution.

Some works have looked at constraints in general ~\cite{Chaudhuri2007,ConstraintBased2005,Dedupalog} but they do not focus on the one-to-one constraint. Su \etal\ do rely on the rarity of duplicates in a given web source~\cite{NegativeExamples} but use it only to generate negative examples.
Jaro's work on linking census data matches records using a Linear Program formulation that enforces one-to-one resolution~\cite{Jaro89}. However no justification or evaluation of the constraint is offered, and the LP formulation can be unnecessarily expensive. The focus of this paper sits squarely on evaluating the benefits of globally constraining entity resolution. We show that while our best-performing one-to-one ER algorithms incur only slightly more computational cost than the unconstrained approach, their statistical performance is significantly greater. We also shed light on the problem of graph matching to maximize precision/recall, demonstrating that it does not simply reduce to max weight graph matching.

Many prior studies have explored resolving entities with mixed-type attributes, by comparing individual attributes to get feature-level scores and then using machine learning to combine these scores into an overall match score~\cite{QDBMUD08-STEM,InfoSys-MultClassSys,Kopcke2010}. A significant number of these studies have considered sophisticated feature-level scoring techniques particularly for strings~\cite{IJCAI-IIW03-Strings} and have compared several standard machine learning combiners including SVMs~\cite{ActiveLearn-KDD02,KDD03-MARLIN,QDBMUD08-STEM,InfoSys-MultClassSys}, decision trees~\cite{ActiveLearn-KDD02,ActiveAtlas-KDD02,KDD03-MARLIN,QDBMUD08-STEM,InfoSys-MultClassSys}, and Na\"ive Bayes~\cite{ActiveLearn-KDD02,InfoSys-MultClassSys}. Our goal in this paper is to develop and analyze effective one-to-one resolution algorithms that are generic in nature and do not rely on feature-level scoring or scoring combiners that are highly tuned using human domain expertise. Thus while our experimental comparison of resolving movie entities involves mixed-type entity attributes, we chose to use very simple feature-level comparisons and a logistic regression combiner that is has been used in entity resolution previously~\cite{JCoopInfoSys,KDD98-logistic,InfoSys-MultClassSys,QDBMUD08-STEM,SIGMOD09-logistic}. Even with an arguably simple approach to scoring we show that entity resolution can efficiently be constrained to produce one-to-one matches.

\section{Generic Entity Resolution} \label{sec:algo}
 
The purpose of this paper is not to investigate alternate scoring functions, but rather to explore generic algorithms for constrained ER.
In this section, we describe the abstract ER problem, the framework for including particular scoring approaches, and several generic algorithms for constrained ER. After describing the generic ER algorithms in this section, we provide specific details of our basic scoring for movie entities in Section~\ref{sec:method}.

We begin with notation of the abstract ER problem. Let \DB{1} and \DB{2} be two \term{databases} each containing a finite number of \term{rows} representing \term{entities}. The dataset sizes need not be the same, and each dataset represents its entities via an encoding in some set $\Domain_i$. The goal of the \term{Entity Resolution Problem} is to identify an unknown target relation or \term{resolution} $\TargetMatch \subseteq \DB{1} \times \DB{2}$, given some information about \TargetMatch; the quality of a retrieved relation $\PredictedMatch\subseteq\DB{1}\times\DB{2}$ is measured by precision and recall. In the exact one-to-one case, the target \term{matching} \TargetMatch\ is one-to-one: for each $x_1\in\DB{1}$ there exists at most one $y\in\DB{2}$ such that $(x_1,y)\in\TargetMatch$, and similarly for each $y_2\in\DB{2}$ there exists at most one $x\in\DB{1}$ such that $(x,y_2)\in\TargetMatch$. 

It is common-place in many-to-many ER to leverage sets of known pairs from \TargetMatch\ and its complement, to learn a scoring function $f:\Domain_1 \times\Domain_2 \to \Reals$ such that \PredictedMatch\ is taken to be all pairs in $\DB{1}\times\DB{2}$ scoring above a threshold $\theta\in\Reals$~\cite{KDD03-MARLIN,SIGMOD09-logistic,QDBMUD08-STEM,Kopcke2010,KDD98-logistic,ActiveLearn-KDD02,JCoopInfoSys,ActiveAtlas-KDD02,InfoSys-MultClassSys}. 

We presume an abstract scoring process composed of several stages: First, to improve runtime performance, we block pairs so that not all pairs in the Cartesian product are scored. After blocking, feature-level scores are generated, which are then combined into an overall score using a supervised learner (logistic regression here). Active learning is employed to reduce the required number of human-labeled training examples. This results in a many-to-many resolution; see Appendix~\ref{sec:generic-scoring} for more details on this abstract scoring process. 

We now describe five generic algorithms for constrained ER. The first two, \AlgMatchMM\ and \AlgMatchFC\ are not one-to-one, but rather serve as baseline comparisons. The remaining three algorithms apply one-to-one constraints.

\subsection{Unconstrained ManyMany}

For unconstrained entity resolution we employ the popular approach \cite{KDD03-MARLIN,SIGMOD09-logistic,QDBMUD08-STEM,Kopcke2010,KDD98-logistic,ActiveLearn-KDD02,JCoopInfoSys,ActiveAtlas-KDD02,InfoSys-MultClassSys} described above which we call \AlgMatchMM: given a score function $f: \Domain_1 \times \Domain_2 \to \Reals$ and threshold $\theta\in\Reals$ simply resolve all tuples with score exceeding the threshold.
The statistical performance of \AlgMatchMM\ relies on score function $f(\cdot)$ to rank pairs in order of likelihood of belonging to \TargetMatch. We shall see that the remaining resolution algorithms exploit structure of \TargetMatch\ to weaken their reliance on a robust $f(\cdot)$. As is common in machine learning and information retrieval, decreasing the threshold $\theta$ trades-off precision for recall. \AlgMatchMM\ takes \bigOh{|\DB{1}|\cdot|\DB{2}|} evaluations of $f(\cdot)$, and the same space.

\subsection{One-to-Many FirstChoice}

As a first step towards globally constrained one-to-one ER, we may consider constraining \AlgMatchMM\ to being one-to-many (or many-to-one). The simplest method of achieving this constraint is to resolve $x_1\in\DB{1}$ with a row of \DB{2} achieving the highest score with $x_1$, provided that this score exceeds threshold $\theta$; if no incident edge is scored higher than $\theta$ then $x_1$ goes unmatched:
\begin{eqnarray*}
\PredictedMatchOM = \left\{(x_1,x_2)\in\DB{1}\times\DB{2} | \forall x\in\DB{2}, f(x_1,x_2)\geq \theta, f(x_1,x)\right\}.
\end{eqnarray*}
Similarly for many-to-one \PredictedMatchMO. This algorithm, which we name \AlgMatchFC, requires the same time complexity as \AlgMatchMM, and a reduced \bigOh{|\DB{1}|} space complexity.

\subsection{One-to-One Entity Resolution}

The added structure of the one-to-one resolution problem suggests that the task be viewed in terms of graph matching. The edge set $\DB{1}\times\DB{2}$, together with the vertices $\DB{1}\cup\DB{2}$ form a complete bipartite graph, weighted by the scores of edges. Target \TargetMatch\ now corresponds to a subset of edges that forms a matching in the graph theoretic sense. Clearly, a resolution \PredictedMatch\ that attempts to approximate \TargetMatch\ should be constrained to be one-to-one, which is in stark contrast with existing approaches in the entity resolution literature.  Assuming that scores \emph{well approximate} the likelihood of a match, retrieving \TargetMatch\ can be seen as selecting edges of high weight.

\textbf{MutualFirstChoice.}
Our first one-to-one meta-\linebreak[4]algorithm, \AlgMatchMFC, is simple and fast to run. It matches two entities iff they prefer each other at least as much as any other entity, with ties broken arbitrarily. This corresponds to running \AlgMatchFC\ from each direction and combining the pairs in agreement, \ie\ taking $\PredictedMatchOM\cap\PredictedMatchMO$. This one-to-one resolution is relatively conservative, and should produce strong results on sparse pair scores.

\textbf{Greedy.}
Our second simple one-to-one algorithm is \linebreak[4] \AlgMatchGr. The meta-algorithm first sorts the tuples by decreasing score, discarding those falling below $\theta$. It then steps through the remaining tuples in order: when a tuple is encountered, where neither of the involved entities have been previously matched, the tuple is included in the resolution.  
The runtime of \AlgMatchGr\ increases by a logarithmic factor in the number of pairs for sorting. The algorithm's space complexity is \bigOh{|\DB{1}|\cdot|\DB{2}|}, while storage of the resolution requires only \bigOh{\min\{|\DB{1}|,|\DB{2}|\}} space.


\textbf{MaxWeight.}
Our third generic one-to-one resolution algorithm, \AlgMatchMW, uses exact max-weight bipartite graph matching. The Hungarian Method or augmenting paths are standard approaches to exact matching, and many approximations exist: indeed \AlgMatchGr\ is a 2-approximation (although in our experiments it achieves significantly better approximations). The running-time of \AlgMatchMW\ is cubic in the maximum database size, and space is the same as \AlgMatchGr.

We view the constrained entity resolution problem as one of graph matching. Intuitively it seems natural to assume that \AlgMatchMW\ should achieve the best performance since it selects pairs with the highest total score. However as we observe in Section~\ref{sec:results} and discuss in Section~\ref{sec:disc} this is not always the case: the target objective of optimizing precision/recall does not generally correspond to maximizing weight.

\section{Experimental Methodology} \label{sec:method}

We implemented the five ER algorithms and applied them to the movies domain. The main sources we used for our experiments were the Internet Movie Database (IMDB)---often used as a primary source of metadata and reviews of movies on the Internet---and Netflix, a popular online rental and streaming service.

We also obtained movie data, both from crawling and querying APIs, from a number of other providers, including  iTunes, AMG and Freebase. We focus our discussion mostly on IMDB and Netflix due to their popularity and large volume of data, while limiting details to proof of concept results on the other sources due to space limitations. In particular we use iTunes and AMG to demonstrate the resilience of constrained resolution using a model trained with IMDB and Netflix only. We also show that our approaches enjoy a large improvement to recall over the crowd-sourced movie entity resolution at Freebase.

\subsection{Movie Datasets}

We accessed the public APIs of both IMDB and Netflix to capture information about as many movies as possible (not including adult titles). Both data sources catalog a large number of movies: over 500k movies from IMDB and over 75k movies from Netflix. Both websites surface unique IDs for their movies, making it easy to distinguish movies within each site.\footnote{For sites without a visible ID, we used the crawled URL. Conceivably the same movie can appear under different URLs, requiring de-duplication; but in the sites used, either the URL was unique or we could find an ID.} 

IMDB generally has more metadata than Netflix, with more alternative titles, alternative runtimes, and a more comprehensive cast list. The Netflix catalog is much smaller than IMDB, as it is focused on movies that are popular as DVD rentals in North America. IMDB, on the other hand has has a much more extensive catalog of foreign movies, those unavailable on DVD, obscure titles, and so forth. However, Netflix is not a subset of IMDB. There are thousands of movies on Netflix that cannot be found on IMDB.

Every movie from the sites we investigated had a title. However, other attributes were not universal. 

As we explain below, the same title can appear in different forms, but it is also worth noting that titles are far from unique, justifying the use of additional features. We can find at least ten movies called ``Avatar". Even in the same year, we observed multiple identically-titled movies (perhaps a form of movie spam). For example, there are several movies ``Journey to the Center of the Earth" from 2008. Figure~\ref{fig:unique-titles} summarizes statistics on replicated titles, and replicated titles with identical release years.

\begin{figure}[t]
\begin{center}
\begin{tabular}{|l|l l|}
\hline
& Non-unique title & Non-unique title \& date \\
\hline
Netflix & 6\% & 0.4\% \\
Imdb & 19\% & 2\% \\
\hline
\end{tabular}
\end{center}
\caption{Replicated movie titles make resolution on title alone infeasible. Surprisingly a non-trivial proportion of distinct movies released in the same year have identical titles.}
\label{fig:unique-titles}
\end{figure}

\subsubsection{Human-Labeled Truth Sets}

In order to train and evaluate algorithms for resolving movies, we gathered several collections of human-labeled truth sets.
\begin{itemize}
\item \textbf{IMDB Top 250.} IMDB maintains a list of the top 250 movies of all time ranked by user-submitted reviews \cite{IMDBTop}. For each entry a human used the Netflix domain search engine and Web search to find an appropriate Netflix match, which was possible for each movie.
\item \textbf{Netflix Top 100.} Like IMDB, Netflix maintains a list of the 100 most popular movies among its subscribers. We followed the same procedure to find 100 known match pairs.
\item \textbf{Random 350.} We selected a uniformly random sample of 350 movies from Netflix, and determined 233 matches within IMDB, noting those 117 movies that had no match.
\item \textbf{Freebase} We used the Freebase API---an open repository of structured data---to find 13,005 movie entities linked to entities in both IMDB and Netflix. Due to the open crowd-sourced nature of Freebase we conjectured that the data would be mostly truthful, while suffering from some benign and perhaps malicious errors.
\item \textbf{Boundary Dataset.} In order to train a machine learning-based score combiner, we collected a set of matches and non-matches that would be difficult to automatically identify. We begun with an initial human-tuned linear scoring function based on the feature-level scores described below. We then took a random sample of 1,234 IMDB-Netflix pairs whose scores were close to 0.5, and manually labeled these. 62\% of the pairs were matches. This approach of training set selection corresponds to active learning~\cite{ActiveLearn-KDD02}.
\end{itemize}

\subsection{Performance Metrics}

We measure the performance of the five entity resolution algorithms via precision and recall. In order to evaluate these metrics, we run each algorithm on the entire IMDB-Netflix blocked set of pairs, and observe non/matched in the resulting \PredictedMatch\ compared to the aforementioned truth sets. Let \TruthPos\ and \TruthNeg\ denote pairs in the truth set $\TruthPos\cup\TruthNeg$ known to be matches in \TargetMatch\ and pairs known not lie in \TargetMatch\ respectively. We measure the following counts
\begin{eqnarray*}
TP &=& |\PredictedMatch\cap\TruthPos|\\
FN &=& |\TruthPos\backslash\PredictedMatch| \\
FP &=& |\PredictedMatch\cap\TruthNeg| \\
&&\ +\ \left|\left\{\left.(x,y)\in\PredictedMatch\right| \exists z\neq y, (x,z)\in\TruthPos\right\}\right| \\
&&\ +\ \left|\left\{\left.(x,y)\in\PredictedMatch\right| \exists z\neq x, (z,y)\in\TruthPos\right\}\right|\enspace.
\end{eqnarray*}
From these \term{precision} corresponds to $TP/(TP+FP)$ while \term{recall} is $TP/(TP+FN)$. Notice that false positives of the second and third kinds are inferred from the truth set. \emph{Due to the movie data being one-to-one, a unique feature of our evaluation methodology is that we infer truth about pairs not explicitly in our truth set.}

\subsection{Feature-Level Scoring}

We retrieved the following attributes to use as features for each movie where available:  titles, release year, runtimes, directors and cast. Space does not permit a detailed description of our feature-level scoring, but our approach is intentionally simple. In brief, our feature scores are:
\begin{itemize}
\item Title: Exact match yields a perfect score. Otherwise, the fraction of normalized words that are in common (slightly discounted). The best score among all available titles is used.
\item Release year: the absolute difference in release years up to a maximum of 30.
\item Runtime: the absolute difference in runtime, up to a maximum of 60 minutes.
\item Cast: a count of the number of matching cast members up to a maximum of five. Matching only on part of a name yields a fractional score.
\item Directors: names are compared, like cast. However, the number matching is divided by the length of the shorter director list.
\end{itemize}
Although the feature-level scores could be improved---\eg\ by weighing title words by TF-IDF, by performing inexact words matches, by understanding common title conventions like <title>:<subtitle>, or that the omission of ``part 4" may not matter while the omission of ``bonus material" is significant---our goal is to show that using the one-to-one constraint alleviates the need to be as elaborate in the development of feature scores. Our approach is intentionally minimalist.

\subsection{Blocking}

In order to avoid scoring all possible pairs of movies---which can grow to a very large number, almost 40b for IMDB-Netflix---we employ the standard technique of blocking (see \eg\ \cite{Kopcke2010}). A key is used to define blocks of movie pairs that are scored. In this case we score pairs whose titles share a normalized non-stopword, thereby producing overlapping blocks. We create an index of normalized non-stopwords for all movies.  To deal with titles that contain no normalized words at all (such as the movie ``+/-"), or only stopwords (such as the movies ``At" and ``3"), we insert special tokens in place of a non-stopword (EMPTY\_\linebreak[4]NORMALIZED\_TITLE and STOPWORDS\_ONLY). We \linebreak[4] only score movies that have an entry in common in the index. Those that do not are presumed to have a low score, which we estimate as zero.

\subsection{Learning How to Combine Scores}

We used the core function for fitting generalized linear models in the R statistical computing environment~\cite{R} for training logistic regression. Before training the model, we first randomly partitioned the boundary dataset into a stratified partition of 856 training and 418 test examples. Each example consisted of the feature-level scores for title, year, runtime, director and cast matches (as described above), and a human label for whether the example pair is truly a match or not. We employed a simple active learning variant of logistic regression: as mentioned above, the boundary dataset was itself generated from examples close to an initial linear model's decision boundary.



\begin{figure}[t]
\begin{center}
\begin{tabular}{|ccccc|c|}
\hline
\textit{Cast} & \textit{Title} & \textit{Year} & \textit{Directors} & \textit{Runtime} & \textit{Intercept} \\
\hline
1.56 & 1.13 & -0.86 & 0.62 & -0.31 & 0.82 \\
\hline
\end{tabular}
\caption{The parameters of the logistic regression model, ordered by decreasing absolute value, fit to the Boundary dataset's training part.}
\label{fig:full-weights}
\end{center}
\end{figure}

The parameter vector resulting from centering, scaling then logistic regression learning is displayed in Figure~\ref{fig:full-weights}. Note that the weights' signs are sensible: small year and runtime \emph{differences} contribute to greater overall score, and are the only negative weights learned. By scaling and centering the data prior to learning, we may compare the relative contributions of the features by comparing the weights' relative magnitudes. Relative to runtime, which is the least predictive feature-level score, the cast's weight is over 5 times larger while the title and year weights are over 3.6 and 2.7 times larger respectively. While it may appear surprising at first glance that cast would be more predictive than movie title, it is clear that the cast is discriminative---even the most prolific stars appear in a limited number of movies, and the combination of stars makes the cast even more unique. This relative weight may also be an artifact of our training set, which included a number of non-matching movies with very high title scores.
In terms of the model's statistical performance, logistic regression achieved a moderate test set accuracy of 89\% under a threshold of 0.5. 



\section{Results} \label{sec:results}

This section presents our experimental results on comparing the five meta-algorithms for entity resolution using a thorough case-study on resolving movies entities, for the movie vertical of a major search engine.

\begin{figure*}[t!]
\begin{center}
\begin{minipage}[t]{\columnwidth}
\centering
\includegraphics[width=7.5cm]{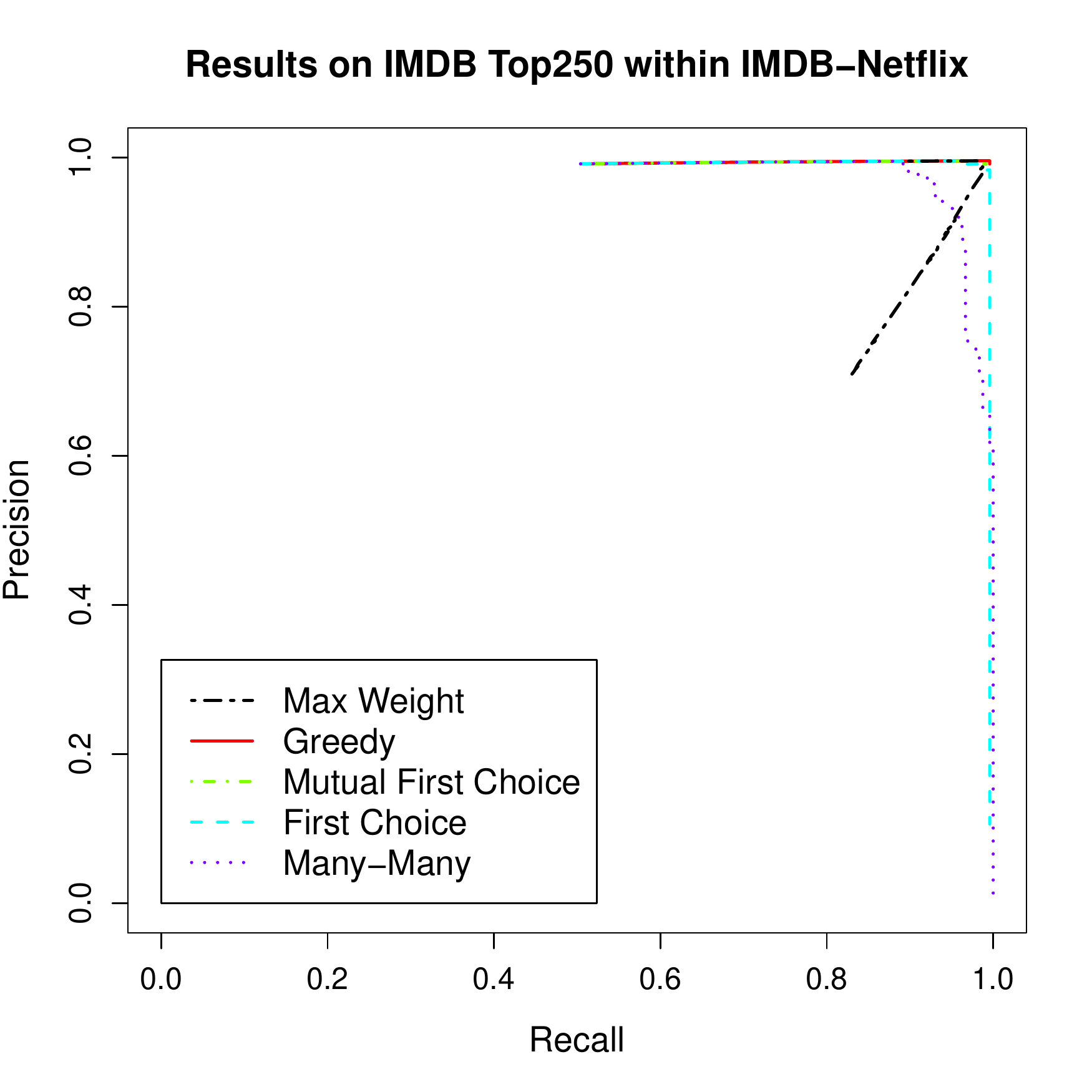}
\vspace{-0.5em}
\caption{Results of testing the entity resolution algorithms on the IMDB Top 250 truth set.}
\label{fig:full-PR-TopIMDB}
\end{minipage}\hfill
\begin{minipage}[t]{\columnwidth}
\centering
\includegraphics[width=7.5cm]{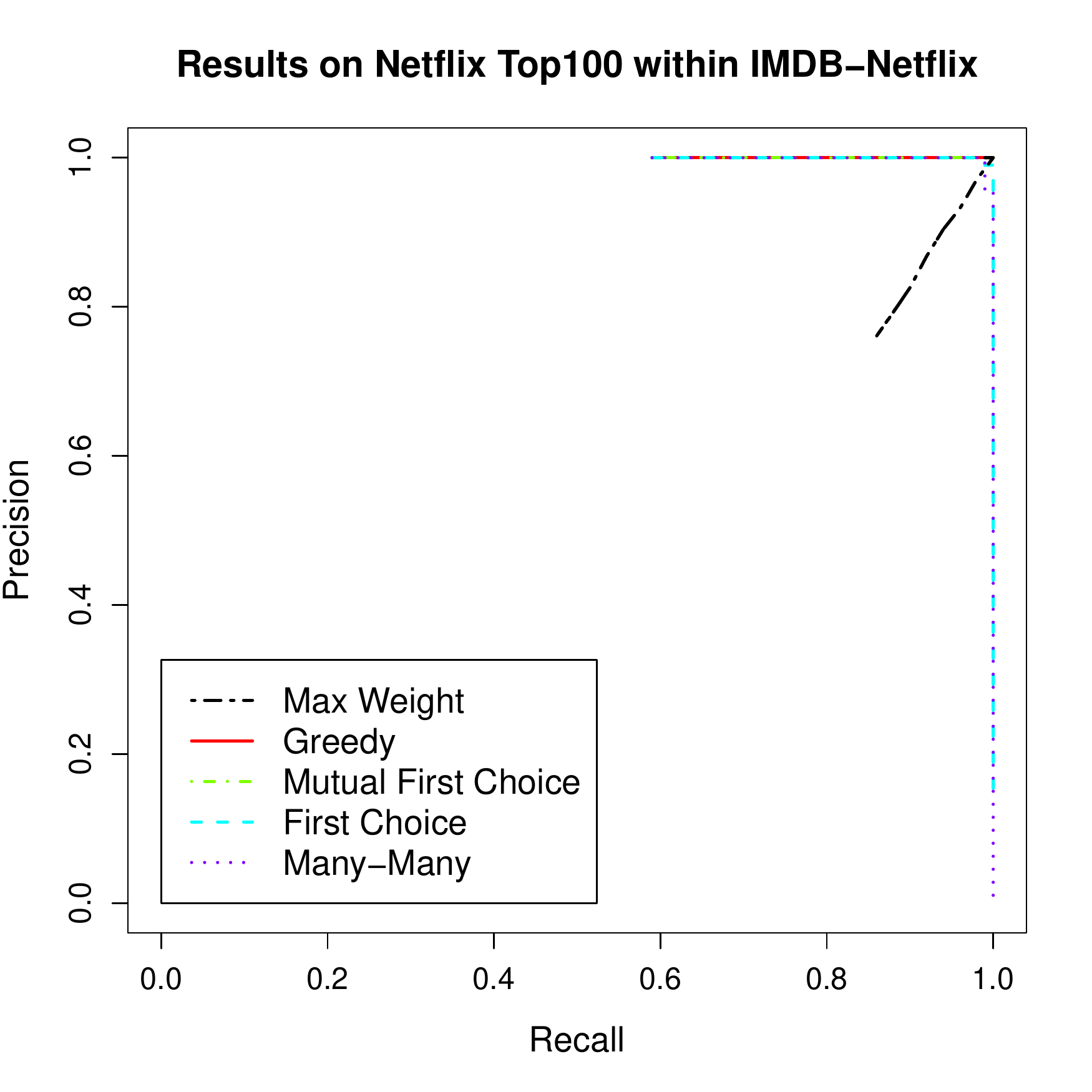}
\vspace{-0.5em}
\caption{Results of testing the entity resolution algorithms on the Netflix Top 100 truth set.}
\label{fig:full-PR-TopNetflix}
\end{minipage}\\
\begin{minipage}[t]{\columnwidth}
\centering
\includegraphics[width=7.5cm]{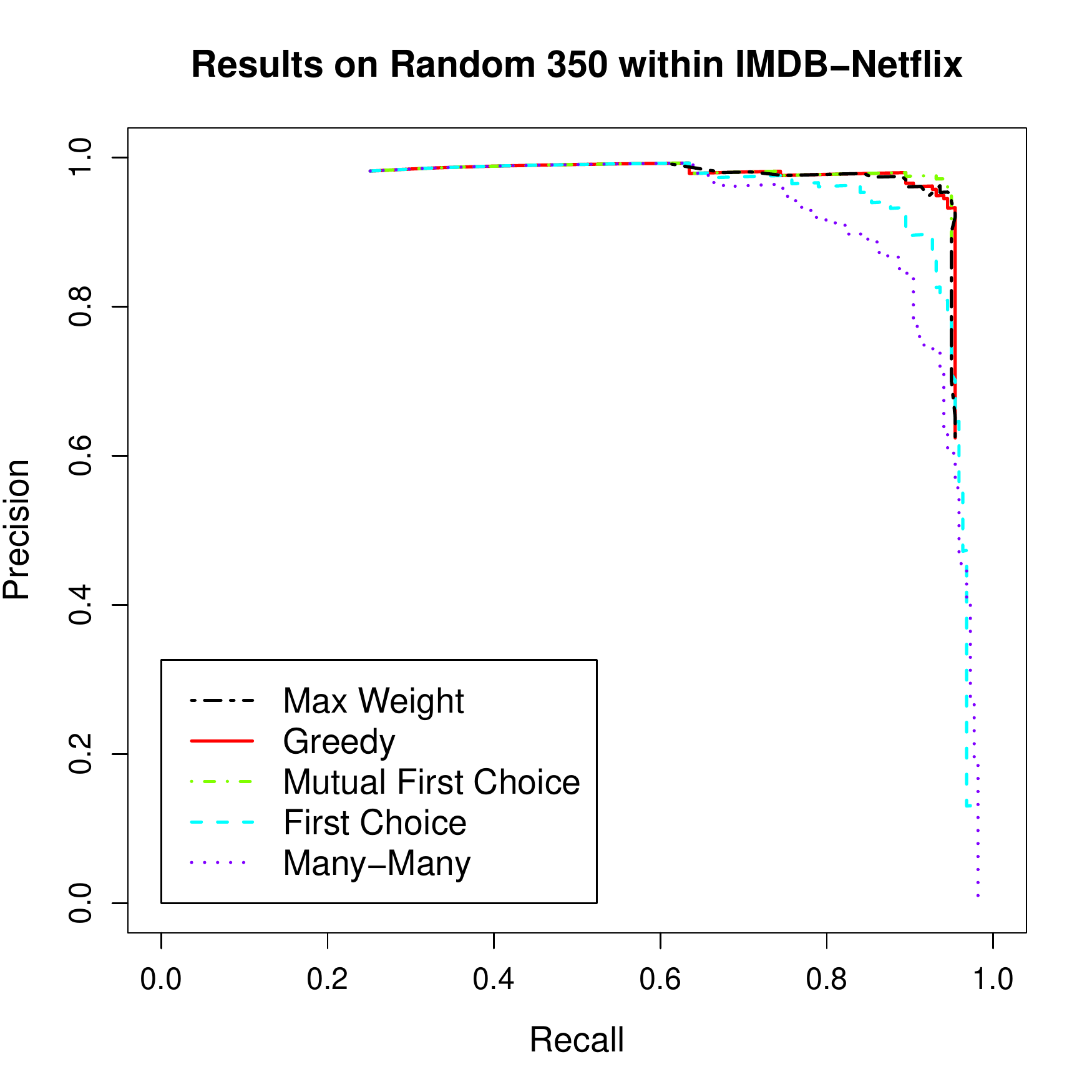}
\vspace{-0.5em}
\caption{Results of testing the entity resolution algorithms on the Random 350 truth set.}
\label{fig:full-PR-Random}
\end{minipage}\hfill
\begin{minipage}[t]{\columnwidth}
\centering
\includegraphics[width=7.5cm]{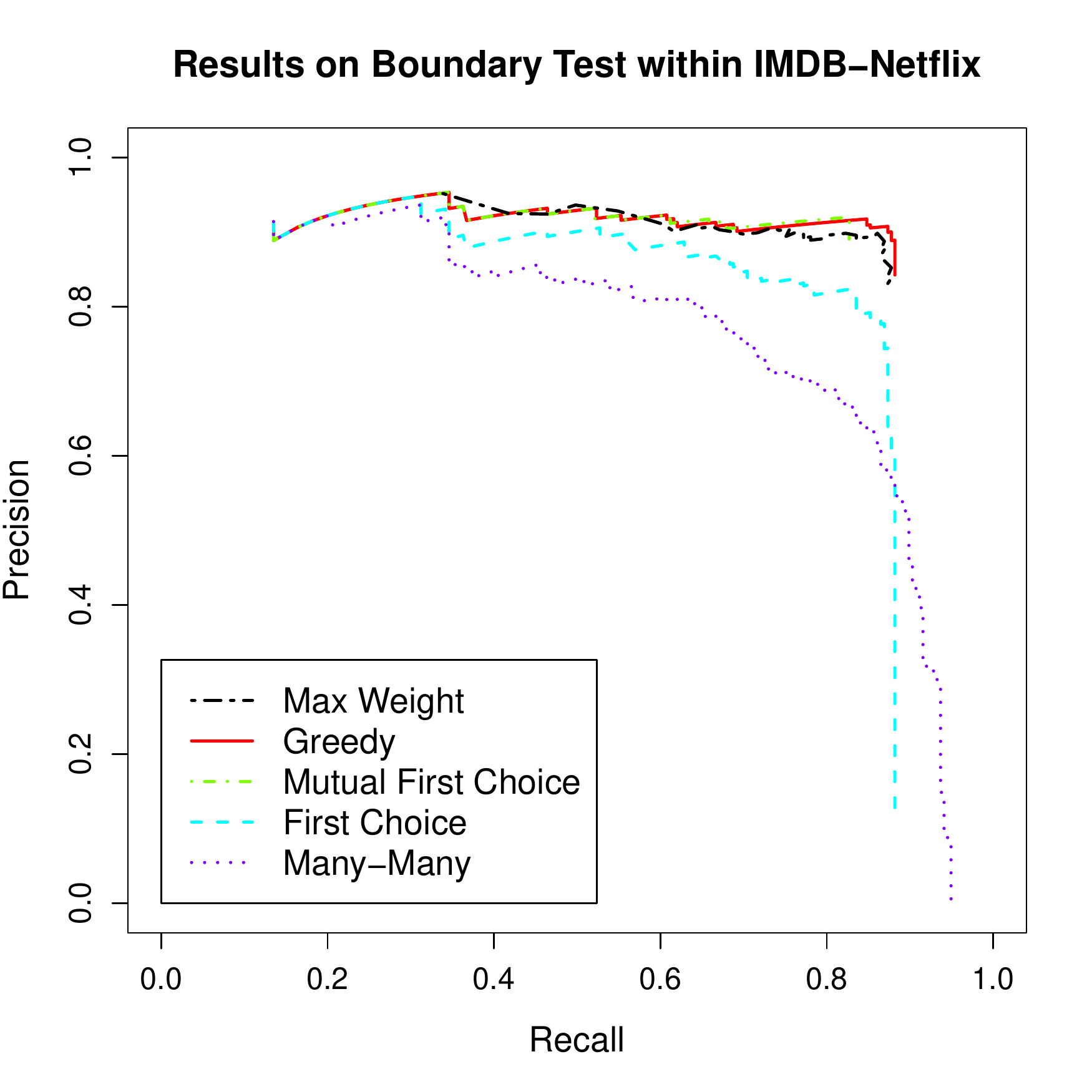}
\vspace{-0.5em}
\caption{Results of testing the entity resolution algorithms on the Boundary dataset's test truth set.}
\label{fig:full-PR-hard}
\end{minipage}
\end{center}
\end{figure*}

\subsection{Validating the One-to-One Assumption}

To assess how close the true IMDB-Netflix matching is to being truly one-to-one, we scored all pairs of movies taken from within each dataset individually. We employed blocking, as two movies with no title word in common are unlikely to be duplicates. We then sorted the results by descending score, and sampled scored pairs at various intervals to get an estimate of the rate of duplicates, which we observed to be declining as the scores decreased, as expected. Based on this observed rate, we conservatively estimate that the true number of duplicates in each source is less than 500. We thus conclude that the IMDB-Netflix movie matching problem is in reality very near one-to-one. The duplicates that do exist are few in number and generally do not get matched to anything, so their impact on our results is not measurable.


\subsection{Comparison of Meta-Algorithms}

We evaluated the unconstrained, one-to-many and one-to-one entity resolution algorithms of Section~\ref{sec:algo}, as described in Section~\ref{sec:method}. Figures~\ref{fig:full-PR-TopIMDB}--\ref{fig:full-PR-hard} display Precision-Recall curves for all five algorithms, on each of the four truth sets from the IMDB Top 250 list, Netflix Top 100 list, a random sample of 350 Netflix movies manually matched with IMDB, and the boundary dataset's test sample.

Viewing the IMDB and Netflix lists of most popular movies as a surrogate for the head of Web or domain-specific movie search queries, we can regard Figures~\ref{fig:full-PR-TopIMDB} and~\ref{fig:full-PR-TopNetflix} as estimating the performance of the respective entity resolution algorithms on head movies. While the two one-to-one algorithms \AlgMatchMFC\ and \AlgMatchGr\ perform almost ideally on both datasets, \AlgMatchMM\ performs almost ideally only on the Netflix Top 100, where movies' Netflix and IMDB representations are complete and therefore easily matched, and fairs worse on the IMDB Top 250. Surprisingly \AlgMatchMW\ performs as expected for only high thresholds receiving strong precision and recall, however the curve doubles back on itself in both figures for decreasing threshold. As we discuss in Section~\ref{sec:disc}, unlike the other one-to-many and one-to-one meta-algorithms for which a decreasing threshold cannot affect previously resolved entities,\footnote{For such algorithms, Precision-Recall curves should be roughly monotonic.} \AlgMatchMW's matching can change drastically even for small decrements to the threshold, potentially producing subtly inappropriate resolutions.

Where the first two figures measure performance on the head, the curves induced by the Random 350 dataset shown in Figure~\ref{fig:full-PR-Random} allow us to compare the meta-algorithms on random movies approximating tail performance. This evaluation tells a far different story: once again the unconstrained \AlgMatchMM\ performs the worst overall, however the one-to-many \AlgMatchFC\ is now significantly worse than the one-to-one matchings whose performance differences are statistical insignificant. Notably \AlgMatchMW's curve behaves like those of the other algorithms; as discussed in Section~\ref{sec:disc} this is due to different sampling used in this third evaluation set. 

Figure~\ref{fig:full-PR-hard} records the performance of the five resolution algorithms on the difficult Boundary dataset's independent test sample. This truth set is made up of examples close to the decision boundary of an initial linear classifier used in the active learning of logistic regression. Thus while we expect this set's examples to be very difficult to match correctly, we see moderate performance from the one-to-one meta-algorithms achieving both precision and recall close to 90\% at the knee, compared to much poorer performances by \AlgMatchFC\ and \AlgMatchMM\ with knee precision and recalls close to 83\%, 81\% and 81\%, 69\% respectively. As in the random set, \AlgMatchMW\ has a curve that behaves like the other one-to-one algorithms.

From these results we conclude that \emph{globally constrained resolution uniformly improves upon unconstrained resolution for the presented large-scale one-to-one movie resolution problem;} that \emph{constrained one-to-one ER is significantly more effective than one-to-many ER;} and that \emph{maximizing matching weight does not necessarily optimize performance of one-to-one ER.}

\pagebreak

\begin{figure*}[t!]
\begin{center}
\begin{minipage}[t]{\columnwidth}
\centering
\includegraphics[width=7.5cm]{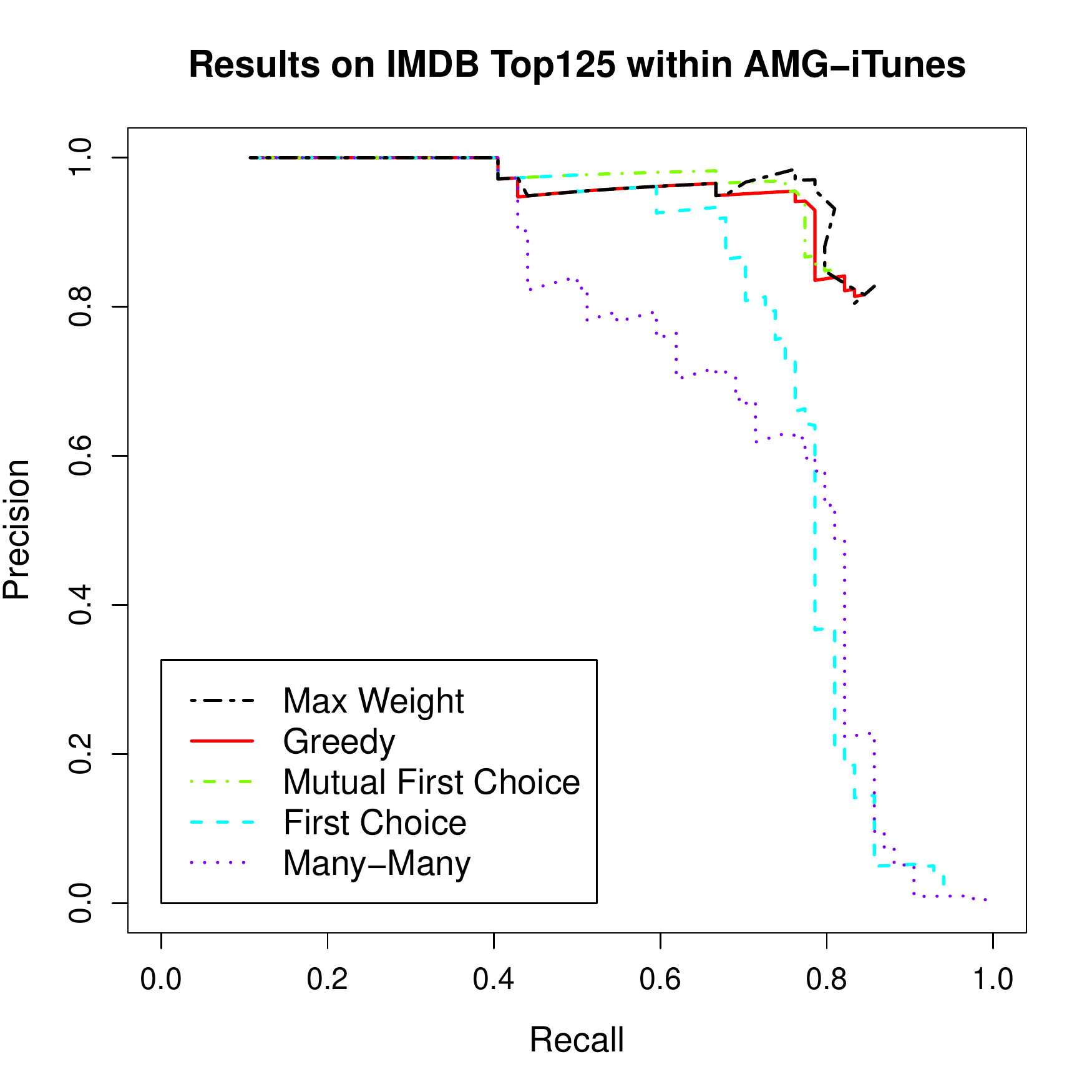}
\vspace{-0.5em}
\caption{Results of testing the entity matching algorithms using IMDB-Netflix trained scores, on a truth set of AMG-iTunes popular movies.}
\label{fig:altData-PR-Top}
\end{minipage}\hfill
\begin{minipage}[t]{\columnwidth}
\centering
\includegraphics[width=7.5cm]{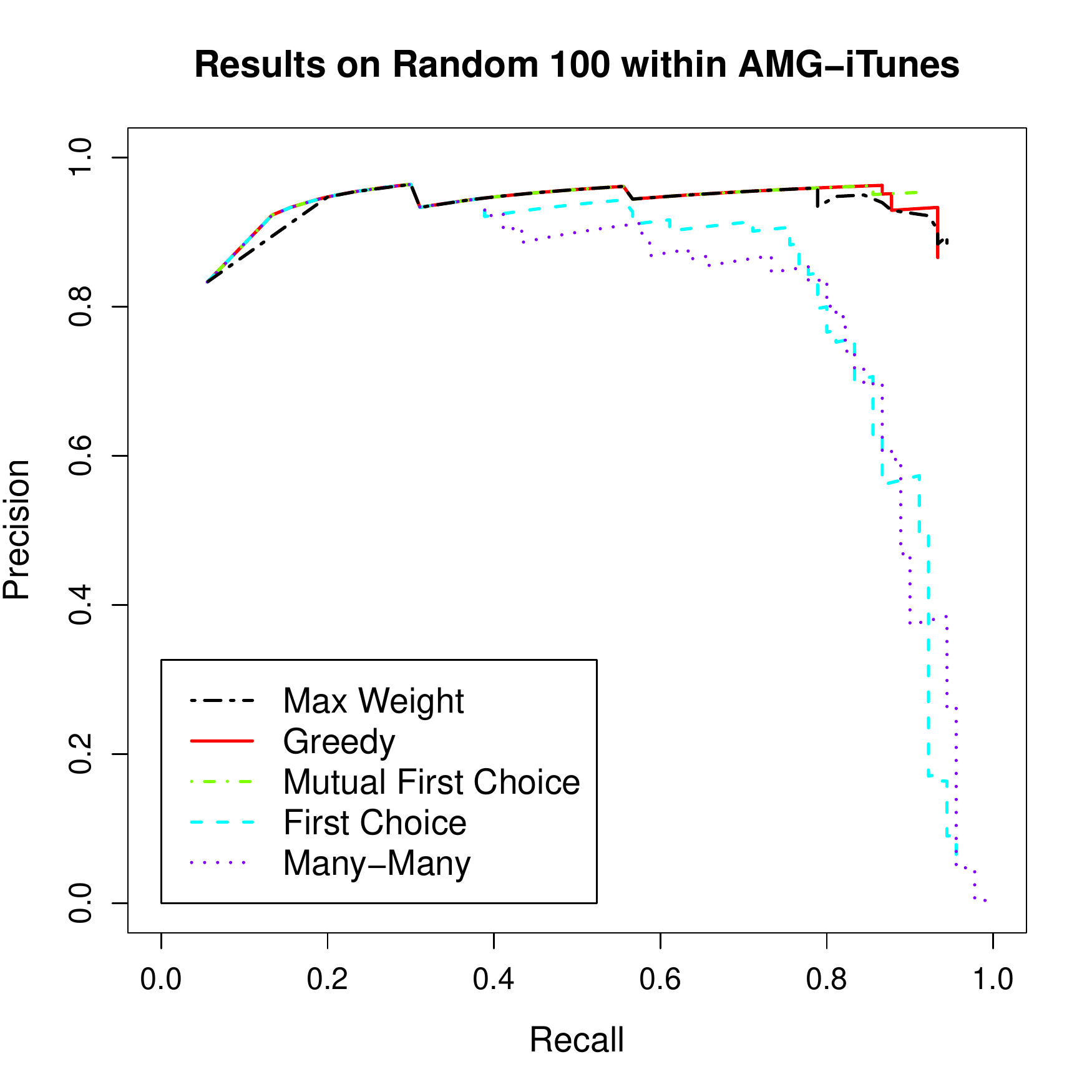}
\vspace{-0.5em}
\caption{Results of testing the entity matching algorithms using IMDB-Netflix trained scores, on a truth set of random AMG-iTunes movies.}
\label{fig:altData-PR-Random}
\end{minipage}\\
\begin{minipage}[t]{\columnwidth}
\centering
\includegraphics[width=7.5cm]{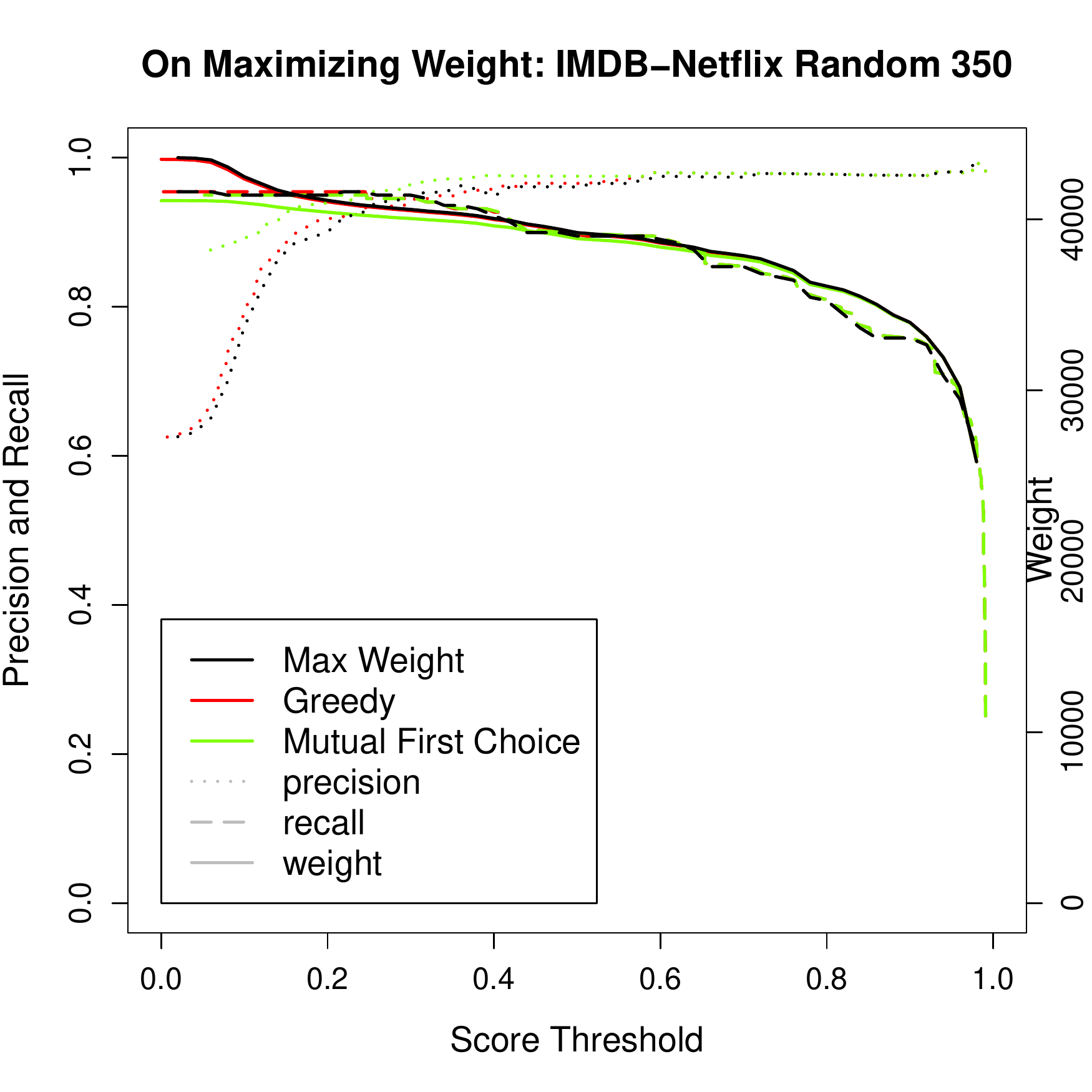}
\caption{All three one-to-one algorithms perform as expected on the IMDb-Netflix random 350, with precision (recall/weight) increasing (decreasing) with threshold.}
\label{fig:maxweight-good}
\end{minipage}\hfill
\begin{minipage}[t]{\columnwidth}
\centering
\includegraphics[width=7.5cm]{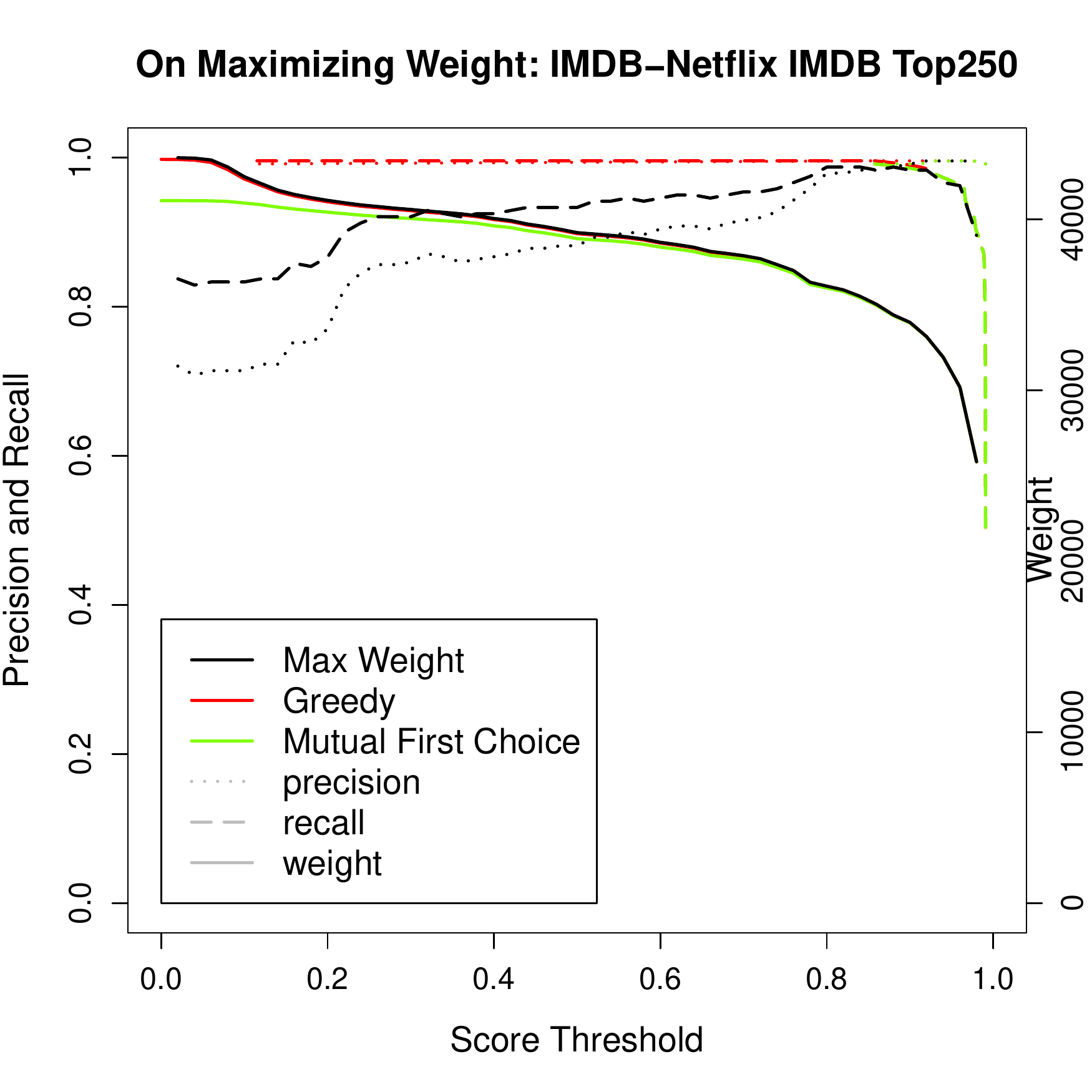}
\caption{Max weight matching's recall is significantly lower than the other one-to-one algorithms for trivial thresholds, and increases with increasing threshold on the IMDb-Netflix IMDb top 250.}
\label{fig:maxweight-bad}
\end{minipage}
\end{center}
\end{figure*}

\begin{figure*}[t]
\begin{center}
\begin{minipage}[t]{\columnwidth}
\centering
\includegraphics[width=\columnwidth]{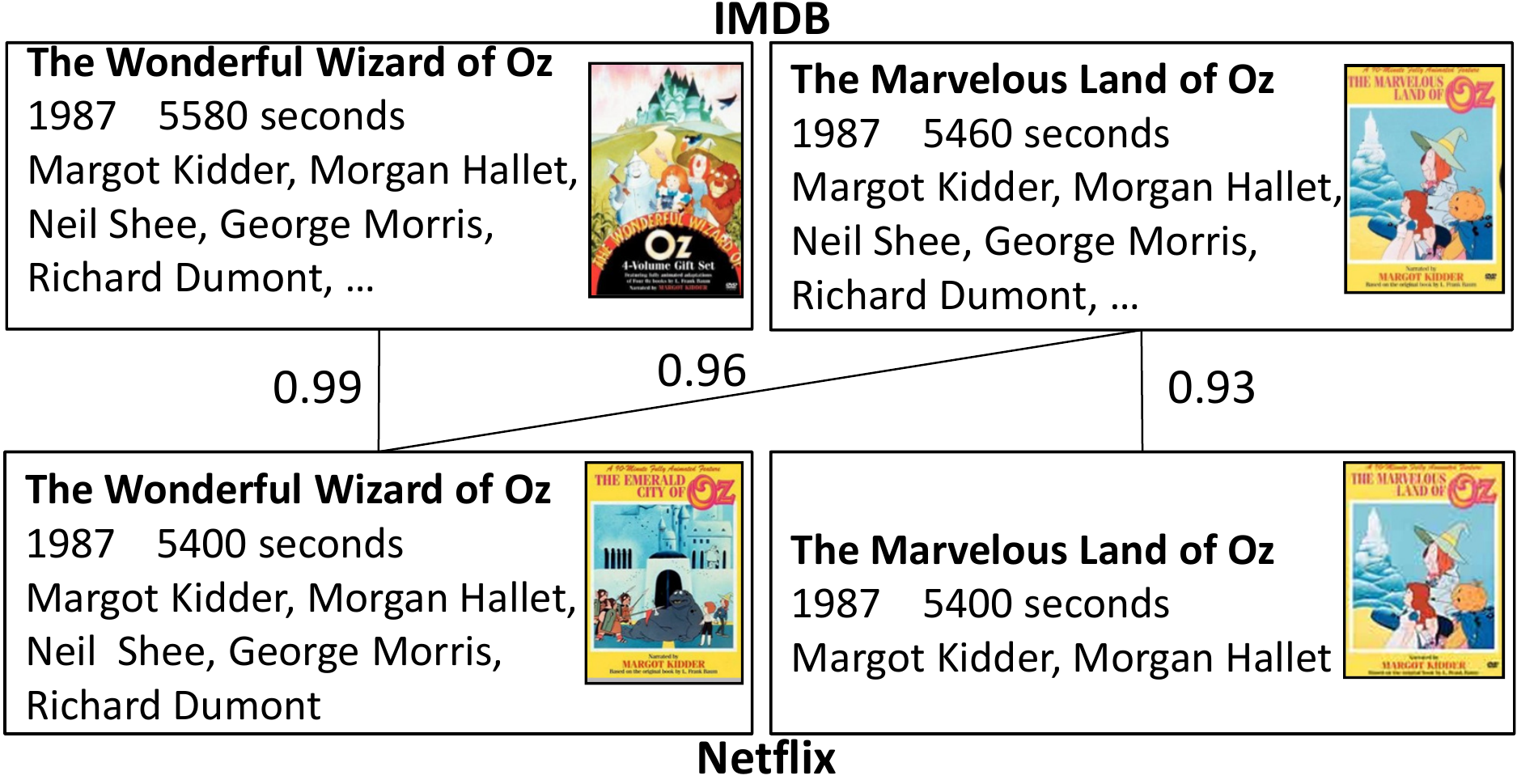}
\caption{Example scores. \AlgMatchFC\ adopts the left two arcs.  \AlgMatchMFC\ adopts only the leftmost arc. \AlgMatchGr\ correctly matches both movies without the false positive.}
\label{fig:discuss-oz}
\end{minipage}\hfill
\begin{minipage}[t]{\columnwidth}
\centering
\includegraphics[width=\columnwidth]{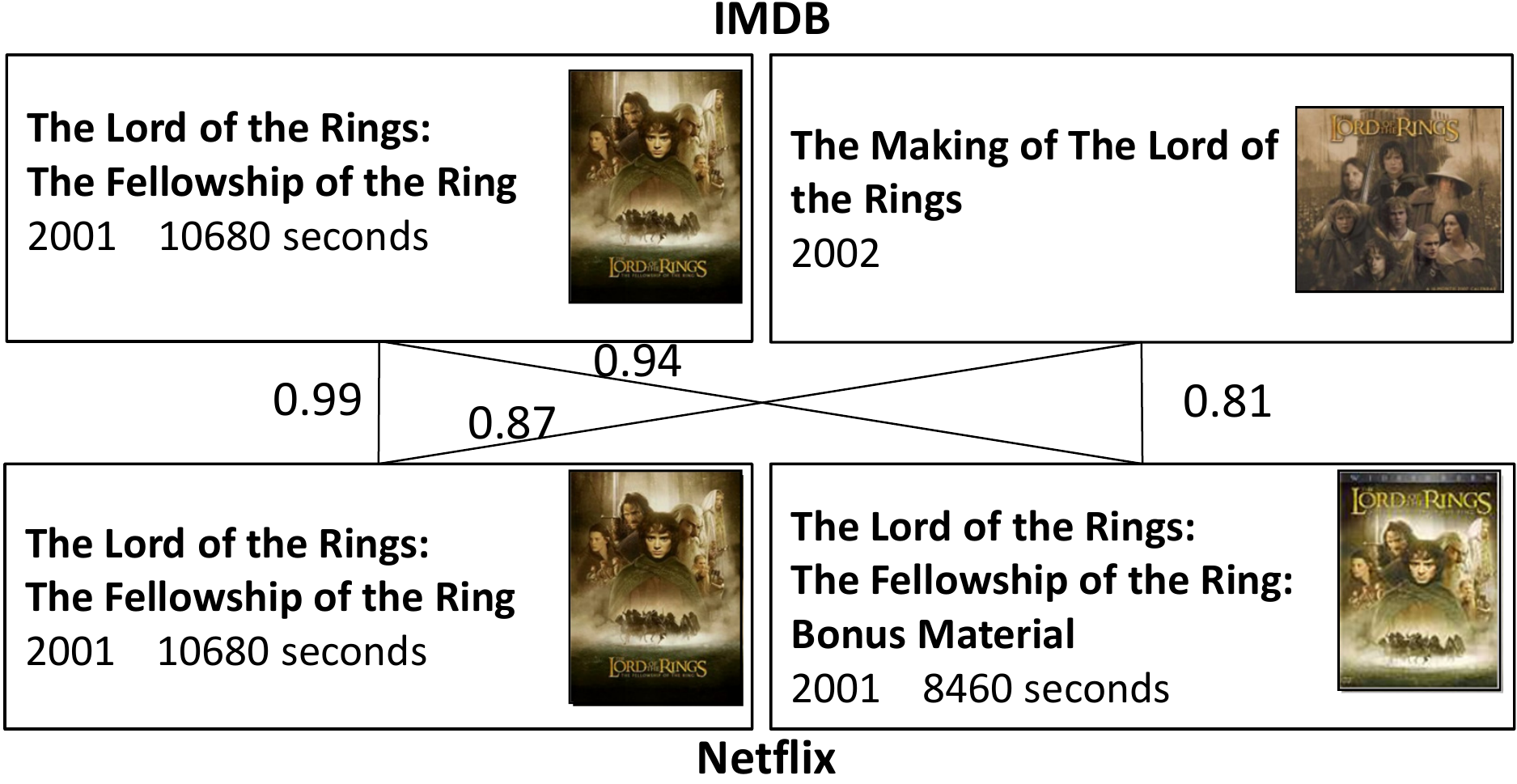}
\caption{Example of scores that can mislead max-weight matching. In this case, max-weight adopts the diagonal arcs. }
\label{fig:discuss-lotr}
\end{minipage}
\end{center}
\end{figure*}

\subsection{Crowd-Sourced Comparison}

We compared our one-to-one resolution with a third party crowd-sourced entity resolution. Freebase is an open repository of structured data that contains many movie entities. We used the Freebase API to download movies and found 13,005 movies that linked to both Netflix and IMDB with IDs contained in our datasets. We compared the matches indicated by Freebase with our constrained resolution \AlgMatchGr, using a threshold of 0.45 which roughly corresponded to the knee's of the Precision-Recall curves above. We found that out of the 13,005 pairs, our algorithm agreed with 12,695 (98\%). We had a human evaluate the 310 additional pairs matched by Freebase and found that 156 (50\%) were incorrectly matched. However, \AlgMatchGr\ matched an additional 29,871 pairs unmatched by Freebase. We took a random sample of 100 pairs from that set and found only one incorrect pair, for a precision of 99\%.
\emph{Compared to Freebase, our algorithm achieved superior precision and significantly improved recall.}

\subsection{Effects of Limited Score Functions}

We conjectured that one-to-one constrained resolution \linebreak[4] would be the most robust to a less accurate score function. To verify our hypothesis, we evaluated the resolution algorithms on new data sources, but without re-training---we use the same score function trained on the IMDB-Netflix Boundary training dataset.  

Figures~\ref{fig:altData-PR-Top} and~\ref{fig:altData-PR-Random} display the results of matching movies between the AMG and the Apple iTunes Store online movie catalogs. The score function is trained on the IMDB-Netflix data to test the resilience of one-to-one entity resolution under poorly tuned scoring. Figure~\ref{fig:altData-PR-Top} shows the results for popular movies and Figure~\ref{fig:altData-PR-Random} shows the result for a random set selected from iTunes manually matched to AMG.  While all algorithms' performances degrade from before, it is clear that \AlgMatchMM\ suffers the most degradation while the three one-to-one algorithms are most robust. This is particularly true in the case of the set of random movie pairs, in which resolution is generally more difficult. This positive result has been repeated with other sources in subsequent work for the production system.

We have found that \emph{the one-to-one constraint can make resolution more robust to the score function.} This implies that practitioners wanting to perform entity resolution can put less effort into the underlying scoring if they make use of one-to-one constraints, simplifying development effort and improving statistical performance.

\section{Discussion} \label{sec:disc}

\subsection{Comparing Constraints by Example}

We have seen that one-to-one constraints can improve aggregate precision/recall, but it is helpful to look in more detail at examples in order to understand the differences between the resolution algorithms. First we can look at how constraints solve problems that occur in the unconstrained case. Consider the movies \emph{Die Hard} and \emph{Die Hard 2}, released in 1988 and 1990 respectively, and with many of the same cast members. With such similar titles, release years and casts, they will have a very high score when compared together. The unconstrained \AlgMatchMM\ algorithm will thus produce four matches (each movie will match with itself and the other) where there are really only two.

The one-to-many constraint is illustrated in Figure~\ref{fig:discuss-oz}, which shows the cartoon movies \emph{The Wonderful Wizard of Oz} and \emph{The Marvelous Land of Oz}, both released in 1987, with the same director, the same cast and a very similar runtime. Note that the diagonal arc has a higher score than the arc on the right because the greater number of matching cast members outweighs the lack of exact title match (the bottom right movie only lists two cast members). 
In this case,  \AlgMatchFC\ would match the Netflix \emph{The Wonderful Wizard of Oz} with both IMDB movies, getting one match right but also one false negative and one false positive.
\AlgMatchMFC\ would only match \emph{The Wonderful Wizard of Oz} between IMDB and Netflix, but not match \emph{The Marvelous Land of Oz} for one false negative. \AlgMatchGr\ would correctly match both movies and ignore the erroneous arc.

\AlgMatchMW\ can be misled by a number of very similar movies. In Figure~\ref{fig:discuss-lotr}, all four movies have similar cast members, years and titles. The ``Making of" and ``Bonus Material" movies are closer in their titles to the full movie than each other, causing the sum of scores on the two diagonal arcs (incorrectly matching them to the full movie) to be  greater than the sum of the vertical arcs (which includes the correct match of the full movie).

\subsection{On the Role of Weight Maximization}

Out of the six comparisons of the one-to-one algorithms' Precision-Recall curves, \AlgMatchMW\ exhibited seemingly unreasonable behavior in two: its curve begins with high precision and recall for high score thresholds, and then doubles back on itself as the threshold decreases. Moreover this occurred only on the IMDB Top 250 and Netflix Top 100 evaluation sets under IMDB-Netflix resolution.

\AlgMatchMW\ may produce drastically different results under small changes in the threshold, while the other one-to-many and one-to-one meta-algorithms' already resolved pairs must go unaffected. Consider Figure~\ref{fig:discuss-lotr}, which shows a pathological case on which \AlgMatchMW\ fails. With a threshold between (0.94,0.99) the 0.99 edge is selected, corresponding to a true positive. When the threshold drops to below 0.87, however, the matching flips to produce two false positives, identifying feature films with one of two distinct bonus material entities. Since the bonus materials share little in common between their titles, their score together is low. Even though they could conceivably be resolved leaving the feature movies to be correctly resolved, the total weight of 1.80 is just shy of the 1.81 weight of the double false positives. Indeed this very situation occurs multiple times for the top movies on IMDB and Netflix, since such movies have associated documentaries and bonus materials cataloged by these two data sources. AMG/iTunes top movies, however, do not contain these bonus materials, perhaps due to a smaller catalog, and so \AlgMatchMW\ achieves a maximum matching by selecting true positives. Finally, the movies contained in the random and boundary sets are not feature films so have no associated satellite movie entities.

Figures~\ref{fig:maxweight-good} and~\ref{fig:maxweight-bad} further explore the interplay between matching weight, precision and recall, for the three one-to-one meta-algorithms. The first figure shows three algorithms that achieve similar precisions, recalls and weights, with these increasing and decreasing respectively with increasing threshold, as is intuitively reasonable. The second figure shows one of the top movie evaluations, with \AlgMatchMW's inferior recall increasing initially due to multiple pathological cases (\ie\ increasing true positives). However we see that for both figures, \AlgMatchMW\ is rarely superior: in the former, \AlgMatchMW's precision converges to best only at threshold 0.5.

Together the pathological case from popular movies, and the results on random tail movies, demonstrate that \emph{weight maximization for entity resolution can fail miserably. What is needed are graph matching algorithms that optimize precision/recall.} 


\section{Conclusions \& Open Problems} \label{sec:conc}

The concept of leveraging socio-economic properties has been fundamental in the history of web search.  We are entering an era in which the Web is becoming increasingly structured, and new socio-economic properties will continue to play an important role. This paper first introduces the property that websites often have an incentive to avoid needless duplication of entities, and then leverages this property by applying one-to-one global constraints to significantly improve ER accuracy and robustness. We evaluate our generic ER framework on a large-scale movie matching task implemented in the movies vertical of a major search engine.

We show that on both head and tail movies, one-to-one resolution is superior to the unconstrained approach. Our many-to-many unconstrained algorithm is representative of other existing approaches: it involves blocking movie pairs to reduce the set of pairs that must be scored, movie attributes are individually scored, and a trained logistic regression model is used to combine these scores. Even with relatively unsophisticated feature-level scores, adding global constraints to the resolution achieves strong performance. We also show that constrained resolution is more resilient to a poorly trained score combiner, comparing the competing approaches by training on one pair of datasets before resolving a second pair of datasets. Finally we compare our one-to-one resolution to a crowd-sourced matching from Freebase, with the result that our algorithm enjoys improved precision and vastly improved recall.

A number of interesting open problems remain for future work. We showed that maximizing matching weight can be a poor surrogate for optimizing precision/recall. What is the link between weight and precision/recall? In particular, can the weights be calibrated so that weight maximization better aligns with our true objectives? Secondly, we have explored generic approaches to constrained resolution that work on top of existing many-to-many algorithms; how can pairs be scored and resolved simultaneously, as opposed to using such a two step process?


\section{Acknowledgments}

We would like to thank Olivier Dabrowski, Craig Hunt, Piali Choudhury and Tristan Lahey for many helpful discussions and assistance relating to this research, Andrew Goldberg for his code used in our implementation of exact max weight graph matching, and the members of MSR Silicon Valley for their help in labeling data.

%
\bibliographystyle{abbrv}
\bibliography{sources}  
%

\appendix

\section{Generic Scoring for ER}\label{sec:generic-scoring}

As described in Section~\ref{sec:algo}, many-to-many entity resolution makes use of blocking to prune down the number of entity pairs to compare, feature-level scores for comparing entities' attributes, and score combination via machine learning~\cite{KDD03-MARLIN,SIGMOD09-logistic,QDBMUD08-STEM,Kopcke2010,KDD98-logistic,ActiveLearn-KDD02,JCoopInfoSys,ActiveAtlas-KDD02,InfoSys-MultClassSys}. This section describes this abstract process, and provides detail on the particular learning algorithm we use for combining feature-level scores.

\textbf{Blocking.}
We adopt a two-step scoring system in which all pairs are first passed through a conservative blocking mechanism $g(x_1,x_2)\in\{0,1\}$ that filters out pairs which are likely to score close to zero---the range corresponds to that of logistic regression which outputs scores $h(\cdot)\in (0,1)$.
We score these pairs zero without explicitly evaluating the 
learned score as is common in other approaches to entity \linebreak[4]
\balancecolumns
\noindent resolution~\cite{Winkler06,Kopcke2010}:
\begin{eqnarray*}
f(x_1,x_2) &=& g(x_1,x_2) \cdot h(x_1,x_2)\enspace.
\end{eqnarray*}
By choosing a mechanism $g$ that is cheap to evaluate, we stand to significantly cut down on the most expensive phase of resolution: scoring. We detail our domain-specific approach to blocking in Section~\ref{sec:method}.

\textbf{Feature-Level Scores.}
We consider the domains of \DB{1} and \DB{2} to be the same. This presumes some normalization during data ingestion, and implies that one dataset may have all null values for a given attribute that it does not support. In these cases scoring $h(\cdot)$ involves first evaluating $d$ feature-level scores $h_1(x_1^1,x_2^1),\ldots,h_d(x_1^d,x_2^d)$, and then combining them using a supervised machine learner~\cite{QDBMUD08-STEM,InfoSys-MultClassSys,Kopcke2010}:
\begin{eqnarray*}
(x_1,x_2) \mapsto h\left(h_1(x_1^1,x_2^1),\ldots,h_d(x_1^d,x_2^d)\right)\enspace.
\end{eqnarray*}
We discuss our specific feature-level scores in Section~\ref{sec:method}, however common choices involve string matching and $L_p$ norms for numeric vectors~\cite{IJCAI-IIW03-Strings,Kopcke2010}.

\textbf{Logistic Regression.}
Logistic regression is a simple but effective method for supervised classification~\cite{Bishop06} used in several previous works on entity resolution to combine feature-level scores~\cite{JCoopInfoSys,KDD98-logistic,InfoSys-MultClassSys,QDBMUD08-STEM,SIGMOD09-logistic}. Let $X$ be a random variable representing the feature-level scores of a pair of entities, and $Y$ be a random Boolean-valued variable indicating whether the pair is a match. Logistic regression employs a parametric model of the posterior likelihood of the form
\begin{eqnarray*}
h(\v{x})\ =\ \Pr{Y=1 \mid X=\v{x}} &=& \frac{1}{1 + \exp(\v{w}\cdot\v{x} + w^0)}\enspace,
\end{eqnarray*}
where $\v{w}\in\Euclidean{d}$ and $w^0\in\Reals$ are the model parameters. One interpretation of logistic regression, is as a result of a generative model under the Na\"ive Bayes assumption (class conditional independence of the features) in which the conditional likelihood functions are Gaussians. An important property of logistic regression, is that the classifier's decision boundary is linear, since a monotonic transformation of the log-odds is linear in the feature vector:
\begin{eqnarray*}
\log_e \frac{\Pr{Y=1\mid X=\v{x}}}{\Pr{Y=0\mid X=\v{x}}} &=& \v{w} \cdot \v{x} + w^0\enspace.
\end{eqnarray*}
To fit the model to a training set of Boolean-labeled feature vectors, the conditional data likelihood---the likelihood of the observed $Y$ values conditioned on the observed features---is maximized via an iterative gradient ascent procedure. Parameter fitting in this way is more robust to incorrect modeling assumptions than Maximum Likelihood Estimation~\cite{MitchellChapter}.

For each entity resolution algorithm, we first train a logistic regression model on a labeled set of entity pairs within and outside \TargetMatch, of moderate size much smaller than $|\DB{1}|\times|\DB{2}|$. With this scoring function $h$ in hand, we may proceed to score all or those unfiltered pairs of entities due to blocking, and continue with the other steps of the particular resolution algorithm. Note that the time complexity of evaluating $h$ on a pair is efficient at \bigTheta{d} if each feature-level score evaluation takes constant time; in reality the constant may be relatively large.


\end{document}